\documentclass[%
reprint,
 amsmath,amssymb,
 aps,
]{revtex4-2}

\usepackage{graphicx}
\usepackage{dcolumn}
\usepackage{bm}
\usepackage{hyperref}
\usepackage[mathlines]{lineno}
\usepackage{xcolor}
\usepackage{tabularx}
\begin{document}

\preprint{NIM-B}

\title{Measurement of the neutron shielding efficacy of magnetite for Proton Therapy Facilities and other applications}
\newcommand*{\HUPCI}{Hampton University Proton Cancer Institute, Hampton, Virginia 23666, USA}
\affiliation{\HUPCI}
\newcommand*{\JLAB}{Thomas Jefferson National Accelerator Facility, Newport News, Virginia 23606, USA}
\affiliation{\JLAB}
\newcommand*{\RAD}{RAD Technology Medical System LLC, Aventura, Florida 33180, USA }
\affiliation{\RAD}

%%%%%%%%%%%%%%%%%%%% authors %%%%%%%%% 
\author {K.~Park} 
\affiliation{\HUPCI}
\author {C.~Keppel, P.~Ambrozewicz} 
\affiliation{\JLAB}
\author {K.~Wright, M.~Kosinski} 
\affiliation{\RAD}

\date{\today}

\begin{abstract}
The neutron shielding properties of high-density concrete and magnetite aggregates were investigated using experimental measurements and Monte Carlo (GEANT4) simulations. Because these materials are widely used in medical accelerator facilities, accurate characterization of their neutron shielding performance is essential for radiation protection and facility design. Experimental measurements were performed at the NASA Space Radiation Laboratory (NSRL) at Brookhaven National Laboratory (BNL) using proton beams with energies relevant to clinical proton therapy. The measured neutron doses were compared with GEANT4 simulations to evaluate the consistency of the simulation methodology with the experimental measurements. Neutron dose attenuation was evaluated for various shielding thicknesses, and the shielding performance of magnetite and conventional concrete was quantified through attenuation coefficients and dose ratios. The results demonstrate that magnetite provides more effective neutron attenuation than conventional concrete for the investigated shielding geometry and proton energies. These measurements provide experimental information for evaluating magnetite as a potential shielding material in medical and research facilities.
\end{abstract}
                            
\maketitle

\section{Introduction}
Ion beam therapy, such as proton therapy for cancer treatment, has been increasingly implemented worldwide due to its advantages, including a lower total dose and the presence of the Bragg peak associated with tissue sparing~\cite{Proton}. The typical proton beam energy used for such medical treatment ranges from 70 to 230 MeV and is mostly generated by either a cyclotron or a synchrotron accelerator.
The construction of proton therapy facilities is significantly more complex and time-consuming than conventional photon-based therapy due to fundamental differences in physics interaction mechanisms, energy deposition, and radiation shielding requirements. One of the major challenges in proton therapy is shielding against neutrons, which are produced as secondary particles from nuclear interactions with protons.  A 250 MeV proton beam produces around 35\% of the neutron ambient dose equivalent per proton dose in Gray (Gy) from direct neutrons, which peaks in energies close to 100 MeV.~\cite{Howell2014}

Neutrons are neutral particles with a mass similar to that of protons and a strong penetration capability, making neutron radiation shielding particularly challenging. Neutrons exhibit a high cross-section for scattering and absorption when interacting with low-Z materials, such as hydrogen-containing objects. As a result, high-density concrete (HDC) is commonly used for neutron shielding in proton therapy facilities. However, the conventional construction with HDC takes many meters of concrete to reduce the high energy neutron enough to optimize these interactions. The molding and curing process of concrete requires several days for stabilization and involves complex procedures to ensure uniformity. This entire process significantly impacts the overall construction timeline and imposes restrictions on building design. Furthermore, the rising cost of HDC is another major concern. There is an ongoing question about whether alternative aggregates can replace the HDC to achieve better neutron shielding efficiency with reduced thickness while also being easier to handle.
 
  A recent study introduced magnetite and hematite ores as potential candidates for neutron shielding~\cite{MDPI}. Since magnetite is more abundant on Earth than hematite, we expect that ore factories can supply it in sufficient quantities at a reasonable cost.
Additionally, a novel construction technique by Rad Technology Medical Systems, LLC offers a modular construction approach~\cite{RADTech}, enabling faster assembly and disassembly. This method utilizes steel structures that can be filled with shielding aggregates in powder form, eliminating the need for curing time and facilitating re-assembly and removal.

The motivation of this study is to evaluate neutron dose attenuation using GEANT4~\cite{GEANT4} and validate the simulations against experimental measurements.
 The measurement of the neutron dose has been carried out with a novel shielding material using the proton beam at the NASA Space Radiation Laboratory (NSRL) at the Brookhaven National Laboratory (BNL)~\cite{NSRL}. The results from both simulations and experiments were then analyzed simultaneously through comparison.

\section{Method}
 The neutron dose for various barrier thicknesses of magnetite shielding material was calculated using Monte Carlo simulations (GEANT4) and then the calculations were compared to experimental measurements. We conducted both methods under the same configuration and compared the results with each other. Concrete was also both measured and simulated for further validation of the procedure and compared to existing global data. The detailed methodology is presented in the following sections.\\
 
\subsection{Simulations}
  GEANT4~\cite{GEANT4} was chosen to estimate the neutron shielding dose beyond the barrier given attenuation through the shielding barrier. We utilized standard CERN libraries ~\cite{GEANT4} of events and physics models for cross-section calculations in GEANT4 and developed custom shielding materials to assess neutron dose attenuation. Simulations were performed with various shielding barrier configurations, considering factors such as thickness, compound ratio, density, and primary beam energy.
The shielding materials were implemented in GEANT4 using their elemental compositions and measured/assumed densities. The concrete used in the GEANT4 simulations was modeled as a homogeneous material composed of H(1.0\%), C(0.01\%), O(52.91\%), Na(1.6\%), Mg(0.2\%), Al(3.39\%), Si(33.7\%), K(1.3\%), Ca(4.4\%), and Fe(1.4\%), with a density of 2.34 g/cm$^3$. 
Magnetite was modeled in GEANT4 as Fe$_3$O$_4$, with elemental mass fractions of approximately Fe = 72.36\% and O = 27.64\%, and a density of 2.88 g/cm$^3$. The material definitions used in the simulation are provided explicitly to allow reproduction of the calculations.

In the simulation, we replicated the experimental setup as closely as possible by incorporating the dimensions of the experimental hall and the major equipment surrounding the detector device. The simulation was conducted for all planned measurement configurations.

\begin{figure}[!htb]
   \begin{center}
	\includegraphics[angle=0,width=75mm,height=70mm]{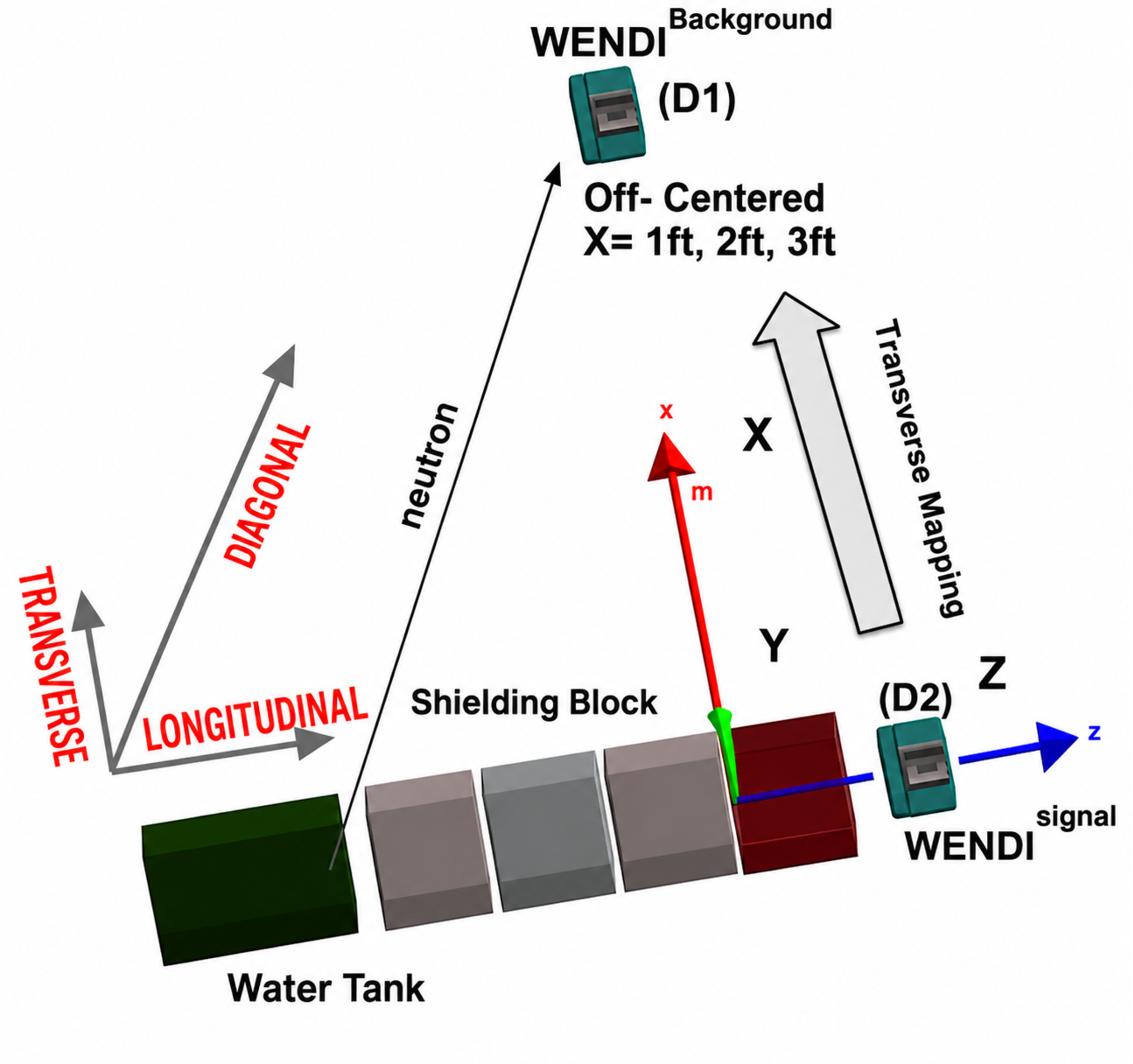}
        \caption[SetUpSim]{
(color online) GEANT4 simulation setup with water tank as target, shielding blocks, and detectors. Measurements were carried out with various detector positions of D1, D2 and numbers of shielding block. 
   }\label{fig:setup_sim}
   \end{center}
 \end{figure}

Figure~\ref{fig:setup_sim} illustrates the simulation setup for the shielding. We defined three coordinate axes: longitudinal (Z), transverse (X), and diagonal (XZ), ensuring accurate application of distances for the inverse square law equation. Shielding blocks were placed directly behind the water tank target with the same beam centering height and attached to its side, minimizing systematic uncertainties in shielding distance determination. The number of shielding blocks directly influenced the interpretation of the results, ensuring consistency in the analysis.

Since the experimental measurement used a finite-volume shielding block, the neutron dose detected after the shielding block was larger than initial Monte Carlo estimates due to significant background noise. The primary source of this background was a large number of neutrons scattered from the target and surrounding objects including the walls.
We defined these scattered neutrons as the main background component, which we aimed to eliminate from the signal detector (D2). To estimate this background, we utilized two identical detectors in both the simulation and the experiment to measure the neutron dose behind the shielding blocks in signal detector (D2) and to simultaneously measure the dose from scattered background neutrons in background detector (D1) outside the shielding block area.

To calculate the neutron dose from the measured counts in the detector, we utilized the relationship between dose, neutron energy, and flux at a virtual detector positioned at the survey point, as described in reference~\cite{MPelliccioni}. We simulated neutrons scattered from the water tank and measured them using the WENDI-II detector through a counting method. By knowing the number of neutrons, their energy, and the geometrical conditions of irradiation, we parameterized the energy dependence and calculated the dose using fluence-to-effective dose conversion coefficients based on an anthropomorphic mathematical model. This model is applicable for neutron energies ranging from 0.0025 eV to 10 TeV. The final dose calculation was determined by averaging the doses from different orientations, including Anterior-to-Posterior (AP), Posterior-to-Anterior (PA), Lateral (Lat), and Iso-symmetric (ISO) configurations. 
 
 To accurately identify direct and indirect background neutron contributions, we specifically constrained the angle between the water tank and detector in the Monte Carlo simulation data by applying the geometrical cut.

\subsection{Experiments}
 We conducted neutron dose measurements using concrete and magnetite shielding blocks at the Brookhaven National Laboratory (BNL). The experiments were performed at the NASA Space Radiation Laboratory (NSRL), which provided multiple proton beam energies of a high beam intensity of approximately $2\times 10^{11}$ per spill, delivered in 4.2-second bunch intervals. This setup allowed us to simulate the annual workload of a proton cancer treatment facility in terms of total beam fluence.
 For our measurements, we requested three proton beam energies, each with the same fluence. The proton beam energy resolution was $\pm 1.8$ MeV for all three energies. However, the beam diameter varied with energy, ranging from 2 to 6 cm. This variation did not affect our results, as the primary goal was to count neutrons rather than measure spatial distribution.
 To generate secondary neutrons, we used a water tank target interacting with the primary proton beam. The water tank had dimensions of $26.7\times31.8\times51.6$ cm$^3$ and it is large enough to fully absorb proton beams up to 250 MeV.\\

 Our measurements were conducted in 2023 (Run 1) and 2024 (Run 2) using various shielding block configurations and proton beam energies. The primary objective of these measurements, was to verify the shielding effectiveness of magnetite as compared to concrete, and critically to validate the simulation results which will provide experimental and simulation benchmarks for evaluating magnetite as a potential neutron shielding material in proton-accelerator applications. Here, concrete data which is well known, is compared to simulation results and previously published data as an overall validation of both magnetite as a shielding solution and the simulation for use in design studies. 
 
 Two shielding materials, concrete and magnetite were used in this experiment. Neutron doses were measured under different shielding configurations by placing 2, 3, 4, and 5 blocks of the material to be evaluated in front of the detector. Each block configuration was tested with three different proton beam energies.
 The concrete data served to evaluate our Monte Carlo simulations via comparison with previous publications and global data. The magnetite data, as a novel shielding material, was used to assess whether our simulations accurately predict its neutron attenuation properties.

Several shielding blocks, each with a volume of 0.028 m$^3$ were prepared for the measurements. The magnetite, in powder form, was placed inside a steel container (0.0127 m thickness), while the concrete blocks were pre-molded with the same volume. Different shielding thicknesses were achieved by varying the number and combination of blocks. The densities of our concrete shielding blocks are ($C1$, $C2$, $C3$, $C4$, $C5$) = (2.201$\pm$0.21, 2.155$\pm$0.22, 2.269$\pm$0.24, 2.376$\pm$0.24, 2.290$\pm$0.22) g/cm$^3$, while five magnetite blocks were prepared using the same-volume container to achieve as uniform a density as possible among the individual blocks. The mass of magnetite placed in each container was determined by weighing the filled container, while the volume was kept constant for all five blocks. The density of each magnetite block was subsequently calculated from its measured mass and the common container volume. This procedure minimized density variations ($<\pm$5\%) among the blocks and provided a consistent magnetite material for the neutron-shielding measurements. We estimated 10\% of uncertainty in density calculation and its average value 2.88$\pm$0.29 g/cm$^3$.

\begin{figure}[!htb]
   \begin{center}
	\includegraphics[angle=0,width=75mm,height=85mm]{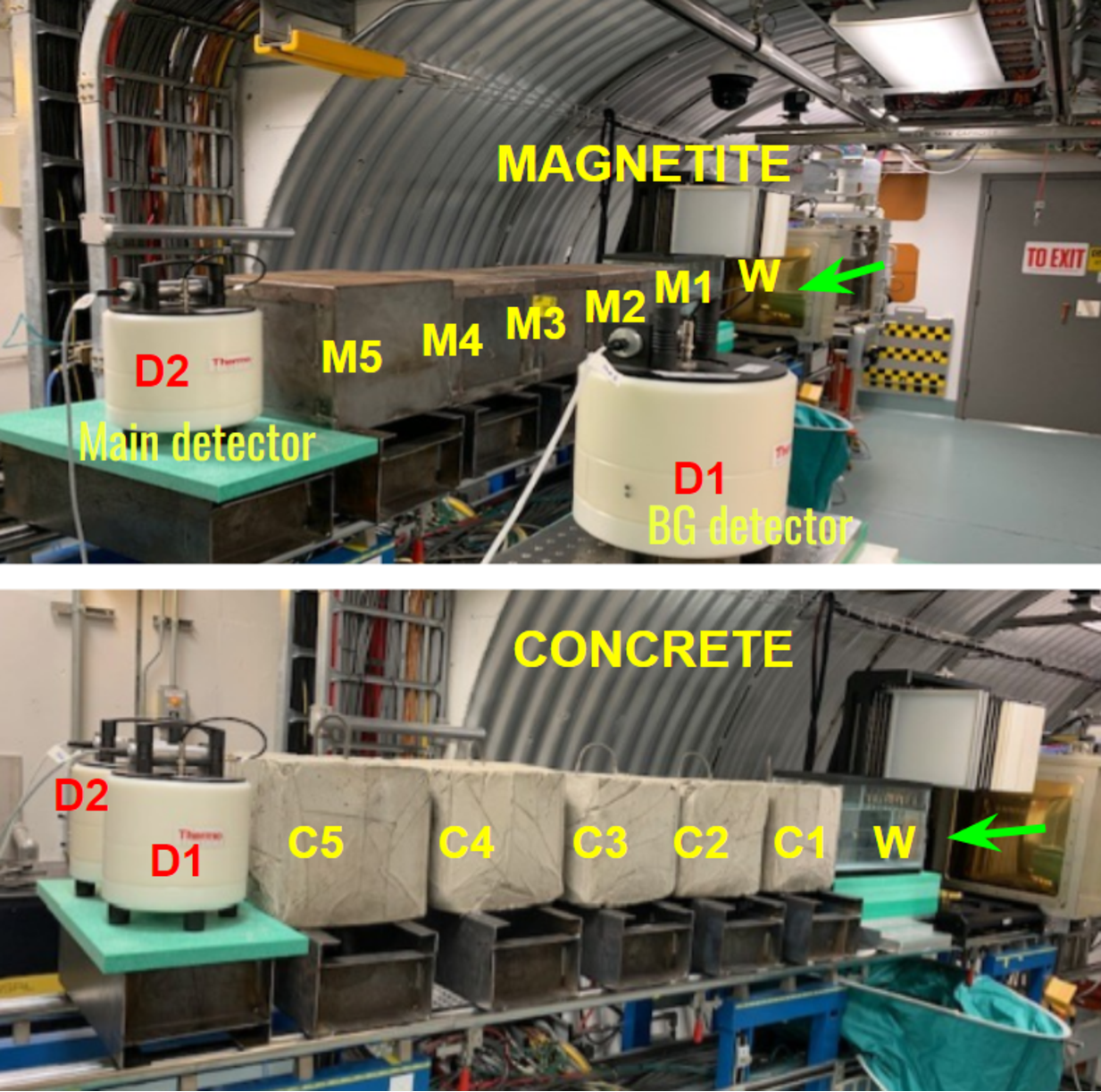}
        \caption[SetUp]{
(color online) Experimental setup with water tank (W) as a target, five shielding blocks, M1...M5 for Magnetite (top) and \& C1...C5 for concrete (bottom) , and two WENDI-II detectors (D1, D2). Green arrow indicates the incoming proton beam direction.
   }\label{fig:setup}
   \end{center}
 \end{figure}

  Figure~\ref{fig:setup} shows the experimental setup with two WENDIs, a water tank, 5 magnetite blocks (top) and 5 concrete blocks (bottom) along the proton beam line at the NSRL experimental hall. To check measurement consistency data taking were done keeping the same configurations (number of blocks, proton energy, detector position) for both concrete and magnetite aggregates.
 
 We used WENDI-II~\cite{WENDI2} neutron detectors, which are widely recognized in the commercial market. The detector was calibrated through the Accredited Dosimetry Calibration Laboratory (ADCL), and its readout system provided dose measurements in $\mu$Sv/min.
To extract neutron dose readings from the WENDI-II detectors, we utilized readout software provided by Laurus Systems. For interfacing between the detectors and the software, we employed a two-channel FHT6020G unit, modifying its firmware to enable simultaneous data acquisition from both detectors. Additionally, a KVM system allowed us to remotely control the readout computer in the radiation environment, where local internet access was unavailable.

Experimental data were collected across various configurations, including different shielding block setups, proton beam energies, shielding materials (concrete and magnetite), and multiple detector positions. In total, we recorded data from over 60 runs, including efficiency measurements taken over 24 hours of beam time. Each run lasted approximately 4–5 minutes, accumulating sufficient statistics of $10^{13}$-$10^{14}$ protons per run, which is comparable to the workload of a clinical proton therapy facility. 

Utilizing two WENDI-II detectors, we estimated the detection efficiency as a systematic uncertainty in dose measurement. To assess this, we tested both detectors under identical conditions, placing them side by side and in an up-and-down configuration to compare their counts. Figure~\ref{fig:setup} (bottom) illustrates one of the measurement setups used for this relative detector efficiency study. We found that the difference in counts measured between the two detectors was less than ±0.5\%.

The shielding blocks were placed behind the water tank, with quantities ranging from 2 to 5 blocks, to evaluate the neutron attenuation effect on detector 2 (D2). Additionally, Detector 1 (D1) was positioned at various transverse distances to assess the impact of random scattered neutrons from the surrounding walls, background. Finally, we conducted measurements at three different proton beam energies (150, 190, and 230 MeV) to analyze the energy dependence of neutron dose.

All collected data were sorted based on the same variable configurations. Each dataset was integrated over the measurement time and then normalized by the total number of protons which was recorded by the beam monitoring system. The final dose value for each detector was recorded in units of $\mu$Sv/spill.
The D2 detector measured the neutron dose after attenuation by the shielding block(s). However, data from D2 contained not only attenuated neutrons from the target but also scattered background neutrons from surrounding objects and the walls of the room. To account for this, we used D1 data to estimate the background contribution to D2.

We provide a more detailed explanation of the background determination which is driven by the scattered neutrons. In general, background is defined as a signal or event that is not of primary interest or does not significantly contribute to the main result. Identifying and quantifying background is a crucial step in the analysis, especially in neutron experiments, where background neutrons are typically present throughout the experimental hall.

In this analysis, D2 serves as the primary device for measuring attenuated neutron doses behind shielding blocks. However, due to the finite size and geometrical limitations of the shielding blocks, they do not fully enclose the detector. As a result, D2 remains partially exposed to scattered neutrons originating from the surrounding structures, which contributes to background contamination in the measurement.

To evaluate this background, we employ D1, which is placed without any shielding blocks in front of it. As such, D1 detects all neutron events based on its position in the hall, including both longitudinal (direct) and transverse (indirect) components originating from the water tank target. However, D1 alone cannot distinguish between these components, since it only records the cumulative neutron fluence.

It is nonetheless necessary to estimate the background contribution to D2, which is located directly behind the shielding block. Owing to its positioning, D2 does not receive direct longitudinal background from the target, but it can receive indirect (scattered) neutrons. Therefore, to correct the D2 measurement, we need to determine the ratio of direct to indirect neutron contributions as captured by D1. To achieve this, we rely on Monte Carlo simulations that allow us to trace the origins and trajectories of detected neutrons. Specifically, we define a neutron kinematic cut based on angular constraints, derived from the geometry—i.e., the tangential longitudinal and transverse distances of D1 from the target. These constraints help us categorize neutron events as either direct or indirect. For the systematic uncertainty study, we applied multiple variations of the kinematic constraints in the background and signal to assess how the direct-to-indirect neutron ratio changes. The resulting spread in the ratio quantifies the uncertainty in the background estimation, which subsequently affects the background subtraction from the D2 measurement.

To quantify the background contribution in D2, we fitted its data to an exponential function as a function of transverse distance from the source. 
The functional form for fit was $D_{\rm BG}=A~\exp(-x/\lambda)+B$, where $D_{\rm BG}(x)$ is the background dose measured at transverse position $x$, $A$ is the amplitude of the distance-dependent component, $\lambda$ is the characteristic transverse attenuation length, and $B$ is a constant background component.

 The exponential form was selected because the measured transverse background decreases approximately exponentially with increasing transverse separation from the direct beam direction. In addition, the stability of the extracted D2 background was tested using an alternative second-order polynomial parameterization. The difference between the resulting background corrections was included as a systematic uncertainty.

The relative contribution of direct and indirect neutrons was investigated as a function of proton beam energy and detector distance. Its functional form is $f_{\rm indirect}= N_{\rm indirect}/ (N_{\rm direct}+N_{\rm indirect})$. For a consistent comparison with the experimental measurements, the total number of incident proton events in each GEANT4 simulation was matched to the corresponding number of protons delivered in each experimental spill.

\begin{figure}[!htb]
   \begin{center}
    \includegraphics[angle=0,width=75mm,height=70mm]{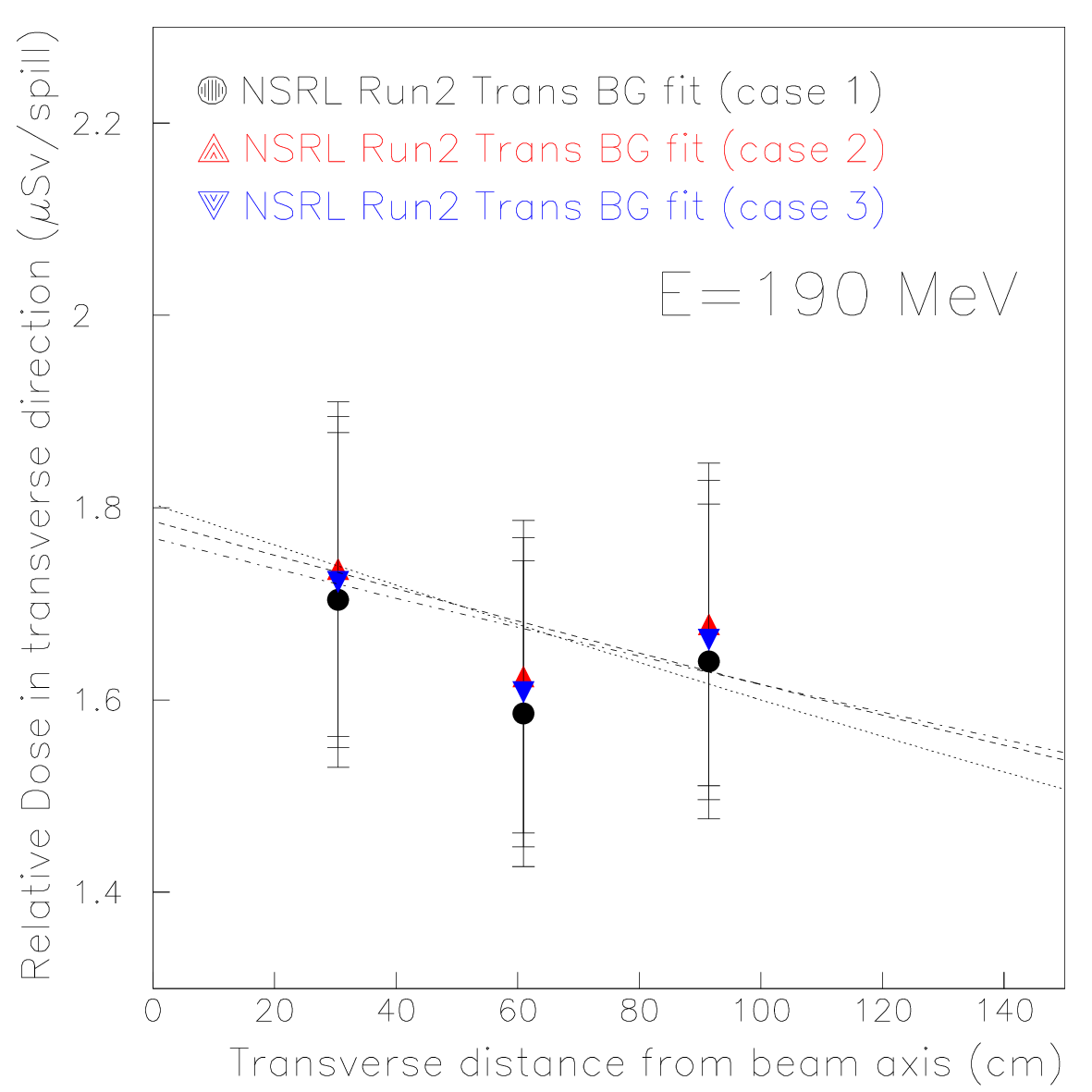}
        \caption[BGFit]{
Transverse neutron dose measured with D1 as a function of the transverse separation between D1 and D2 for 190 MeV protons. Cases 1, 2, and 3 correspond to the three-, four-, and five-block shielding configurations, respectively. The curves represent the exponential parameterizations used to estimate the indirect scattered-neutron contribution at the D2 position.
   }\label{fig:BGfit}
   \end{center}
 \end{figure}

We then applied an indirect neutron fraction derived from the simulation to distinguish direct and indirect scattered neutron events from water tank detected by D1. This process was performed for all measurement configurations by fitting the data accordingly. Figure~\ref{fig:BGfit} presents an example of the transverse background data fit using an exponential function. The transverse-distance-dependent dose represents neutrons reaching the detector through side and back scattering. A residual scattered contribution at 0 cm was estimated by extrapolation of the transverse-distance dependence using the D1 measurement-to-simulation ratio. Three transverse distances data show from different longitudinal distance data sets (case 1,2,3) which are with three, four, and five blocks configuration. \\

 The simulated indirect neutron fraction ranged from 0.33 to 0.42 (33-42\%) over the investigated proton energies and detector positions. At 150 MeV, the values were 0.35, 0.36, 0.36, and 0.42 at distances of 90, 105, 120, and 160 cm, respectively. At 190 MeV, the corresponding values were 0.34, 0.36, 0.37, and 0.41, while at 230 MeV they were 0.33, 0.34, 0.34, and 0.39. These results quantify the contribution of indirect neutrons to the total simulated neutron fluence and were included in the analysis.
This approach allowed us to isolate the indirectly scattered neutron background detected by D1. We then fitted this background data and applied the extrapolated background contribution to D2. Once the background at the D2 position was determined, we subtracted it from the measured dose at D2 to obtain the corrected neutron dose from the target.

\begin{figure}[!htb]
   \begin{center}
     \includegraphics[angle=0,width=75mm,height=70mm]{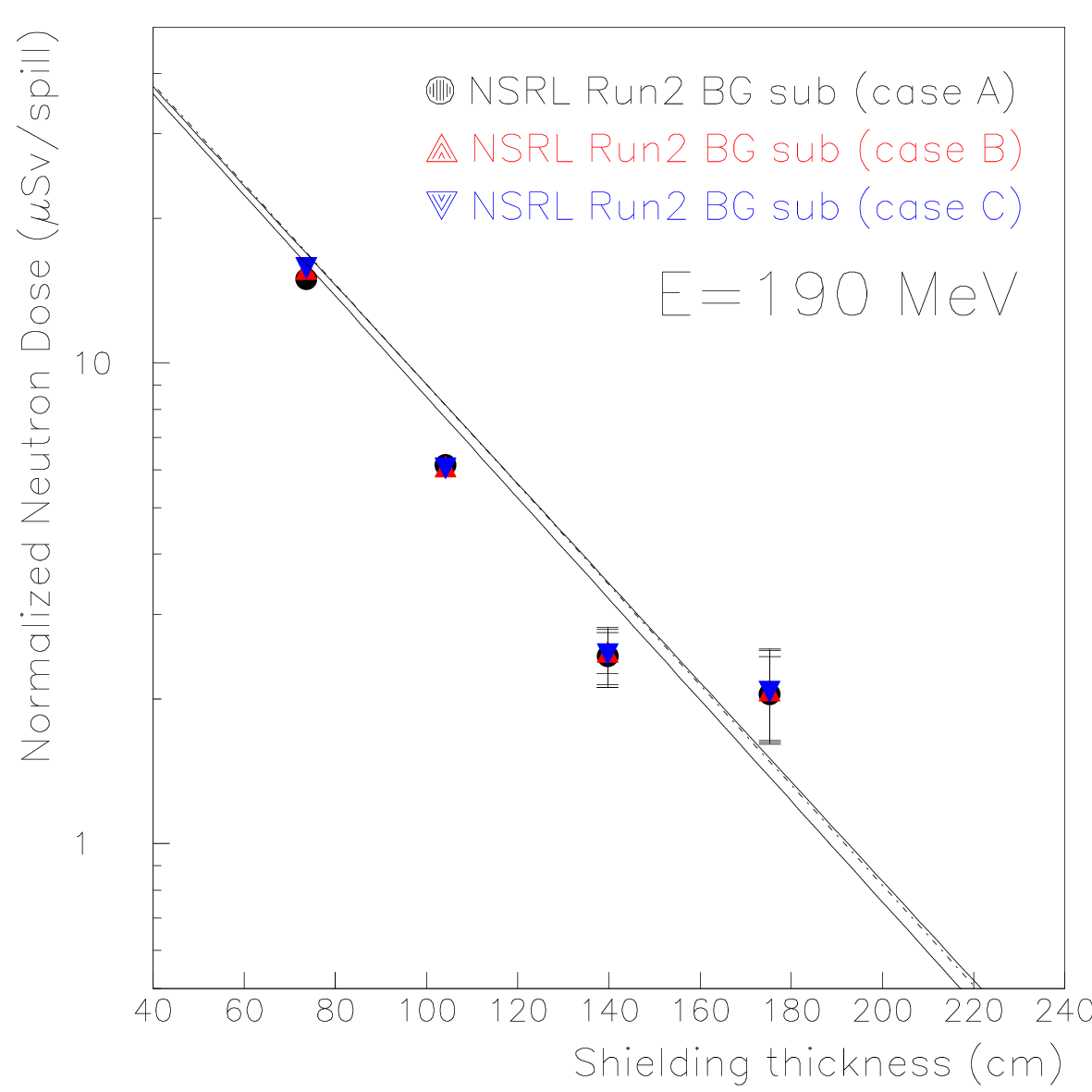}
        \caption[Results]{
Corrected neutron dose measured by D2 as a function of shielding distance for 190 MeV protons. Cases A, B, and C correspond to the background corrections obtained from Cases 1, 2, and 3 in Figure 3, respectively. The spread among the curves represents the systematic variation associated with the background-extraction procedure. Block 1 measurements were not performed because of the high expected neutron dose and the associated risk of exceeding the detector's operational range.
   }\label{fig:BGsubtracted}
   \end{center}
 \end{figure}

We extracted fit parameters for measurement data that were various shielding block configurations and three different proton beam energies. Figure~\ref{fig:BGsubtracted} presents one of the three proton beam energy results after background subtraction. Measurements for Block 1 were not performed because the expected neutron dose was sufficiently high that irradiation could exceed the detector's operational range and potentially damage the detector. Therefore, Block 1 was excluded from the measurement for detector safety. The data show a normalized neutron dose per spill as a function of shielding distance. Each data point shows after applying background subtraction using three different transverse distance data fits (case A,B,C) which are 1, 2, and 3 from figure~\ref{fig:BGfit}. To assess systematic variations, the exponential function ($D_{\rm BG}(x)=A \exp(-x/\lambda)+B$) was used as the nominal parameterization because the measured transverse background showed an approximately exponential decrease with distance. The second-order polynomial ($D_{\rm BG}(x)=A x^2 +B x +C$) was used as an independent alternative to evaluate the sensitivity of the extracted D2 background to the functional form. The difference between the two resulting background corrections was included as a systematic uncertainty.

\section{Results}
%\

In this section, we present not only our measurements and simulations but also the validation by comparing them with published world data. Our validation process follows three key steps:
\begin{itemize}
    \item[A.] Comparison of neutron dose for concrete shielding between GEANT4 simulations and experimental measurements at NSRL.
    \item[B.] Comparison of concrete shielding data from this study with existing world data.
    \item[C.] Comparison of magnetite shielding data between our GEANT4 simulations and experimental measurements.
\end{itemize}

Finally, we discuss key insights gained from our simulation results and evaluate the reliability of our simulation methodology. Since experimental data for magnetite as a neutron shielding material is scarce, we evaluated our simulation and experimental results using concrete, which is well-documented in existing literature. Our concrete density was approximately 2.16 - 2.35 g/cm$^3$, so we selected world data publications that featured shielding materials of similar density. However, most of the world data we referenced were obtained with different neutron sources, such as a 50 MeV proton injector with an iron solid target or a PuBe neutron source from a nuclear reactor.

\subsection{Comparison of GEANT4 simulations with experimental concrete-shielding measurements}
We conducted simulations using various primary proton beam energies of 150, 190, and 230 MeV, with a water tank as the target and various concrete shielding thicknesses. For validation, we compared the neutron dose behind various thick concrete barriers in our GEANT4 simulations with experimental data from NSRL-Run 2, where we used five shielding blocks (each approximately 0.028 m$^3$ in volume).
Table~\ref{tab:5feet} presents the neutron dose evaluated at D2 with a 1.524 m shielding barrier in the simulation, compared to the measured dose using five blocks in the NSRL-Run 2 experiment. The results indicate an approximately 40\% relative difference in dose measurements. The total experimental distance for the five-block configuration was 1.753 m, whereas the corresponding GEANT4 geometry used a nominal shielding thickness of 1.524 m. This difference is attributed to air gaps between adjacent concrete blocks and imperfect block alignment resulting from their imperfect shapes and physical alignment in the experimental setup. The associated uncertainty arising from the inter-block air gaps was evaluated and included in the systematic uncertainty budget.

 We identified the major sources of uncertainties which are background identification ($\sigma_{\rm BG}$), functional form used for background fitting ($\sigma_{\rm fit}$), concrete density ($\sigma_{\rho}$), neutron fluence-to-dose conversion coefficient ($\sigma_{\rm conv}$), and the rest of detector efficiency, proton beam energy, shielding distance/geometry ($\sigma_{\rm other}$).
These yield the principal systematic uncertainties as approximately 4–11\% for background identification, 5\% for functional form, 1–6\% for concrete density, and 3\% for the conversion coefficient, with a total of 7.1–13.8\%.
We have clarified how these uncertainties affect the final neutron dose. For independent contributions, the total systematic uncertainty is calculated as quadrature sum of each uncertainty, which is shown in $\sigma_{\rm sys}= \sqrt{\sigma_{\rm \rm BG}^2+\sigma_{fit}^2+\sigma_{\rho}^2+\sigma_{\rm conv}^2+\sigma_{\rm other}^2}$.

\begin{table*}[!htb]
\centering%% 
\caption{Neutron dose comparison in 5-block shielding thickness barrier between GEANT4 and NSRL-Run 2 measurement for a proton energy E = 230 MeV. A block is 0.0283 m$^3$ volume.\\}\label{tab:5feet}
\renewcommand{\arraystretch}{1.2} 
\begin{tabular}{c|c|c|c|c}%%
\hline
  Configuration & Total thickness (m) & Total mass (kg) & Average density (g/cm$^3$) & Est. Dose (mSv/yr)+Stat.+Sys.\\ %% 
\hline
\hline
NSRL-Run 2, 5 blocks & 1.753 & 367.3 & 2.255 & 0.9 $\pm$ 0.02 $\pm$ 0.015 \\
GEANT4, 5 blocks & 1.524 & 332.4 & 2.340 & 0.5 $\pm$ 0.03 (stat)   \\
\hline
\end{tabular}
\end{table*}

The five-block configuration has a nominal geometric thickness: $t=5\times (0.3048)$ m $= 1.524$ m. 
The experimental concrete density range reported in the manuscript is 2.16 - 2.35 g/cm$^3$.
Therefore, for a 1.524 m barrier, the corresponding mass thickness range is $= 2.160\times 152.4 = 329.3$ g/cm$^2$ to  $= 2.350\times 152.4 = 357.78$ g/cm$^2$.
For the simulation density of $2.34$ g/cm$^3$, $= 2.34\times 152.4 = 356.6$ g/cm$^2$.
Thus, the simulation corresponds closely to the high-density end of the measured concrete range. For a block volume of approximately 0.0283 m$^3$, the mass range corresponding to $2.16-2.35$ g/cm$^3$ is approximately $m=61.1-66.5$ kg/block, or approximately 305.5-332.5 kg for five blocks. 

However the experimental five-block stack contains inter-block air gaps and has a measured overall length of 1.753 m, whereas the GEANT4 model uses an idealized continuous concrete thickness of 1.524 m, the dose values in Table I should not be interpreted as a comparison at identical physical shielding thickness.

\begin{figure*}[!htb]
   \begin{center}
	\includegraphics[angle=0,width=58mm,height=68mm]{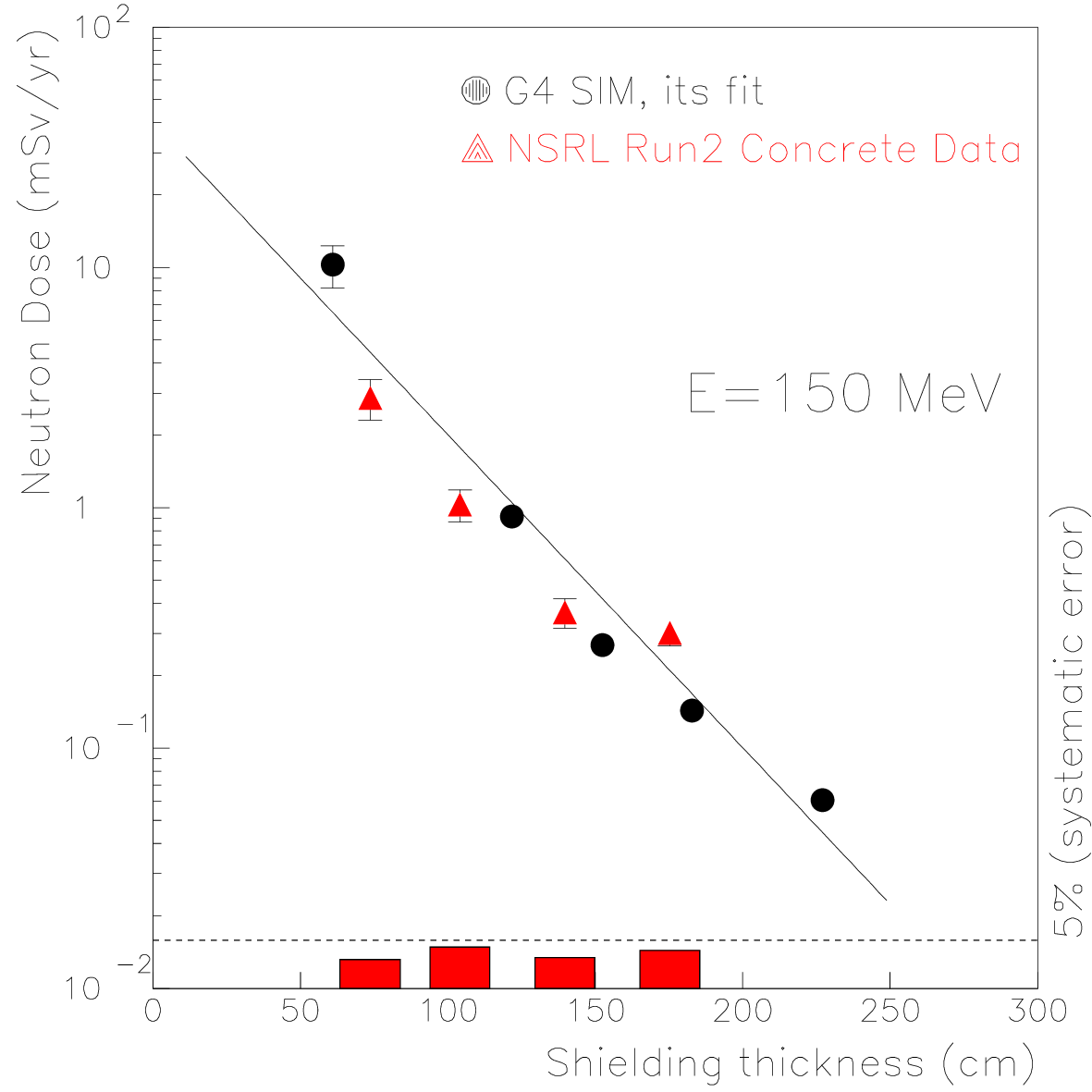}  
    \includegraphics[angle=0,width=58mm,height=68mm]{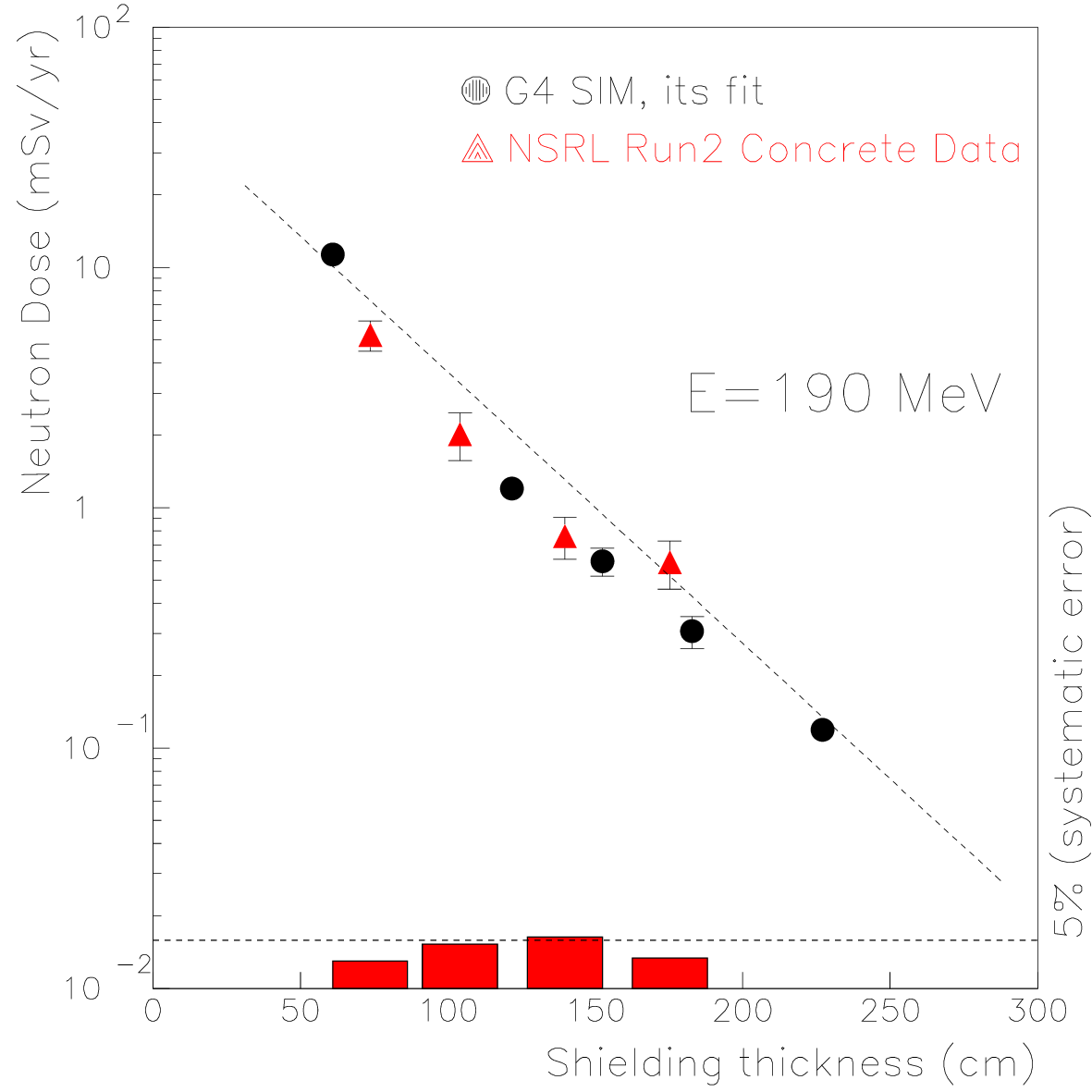}
   	\includegraphics[angle=0,width=58mm,height=68mm]{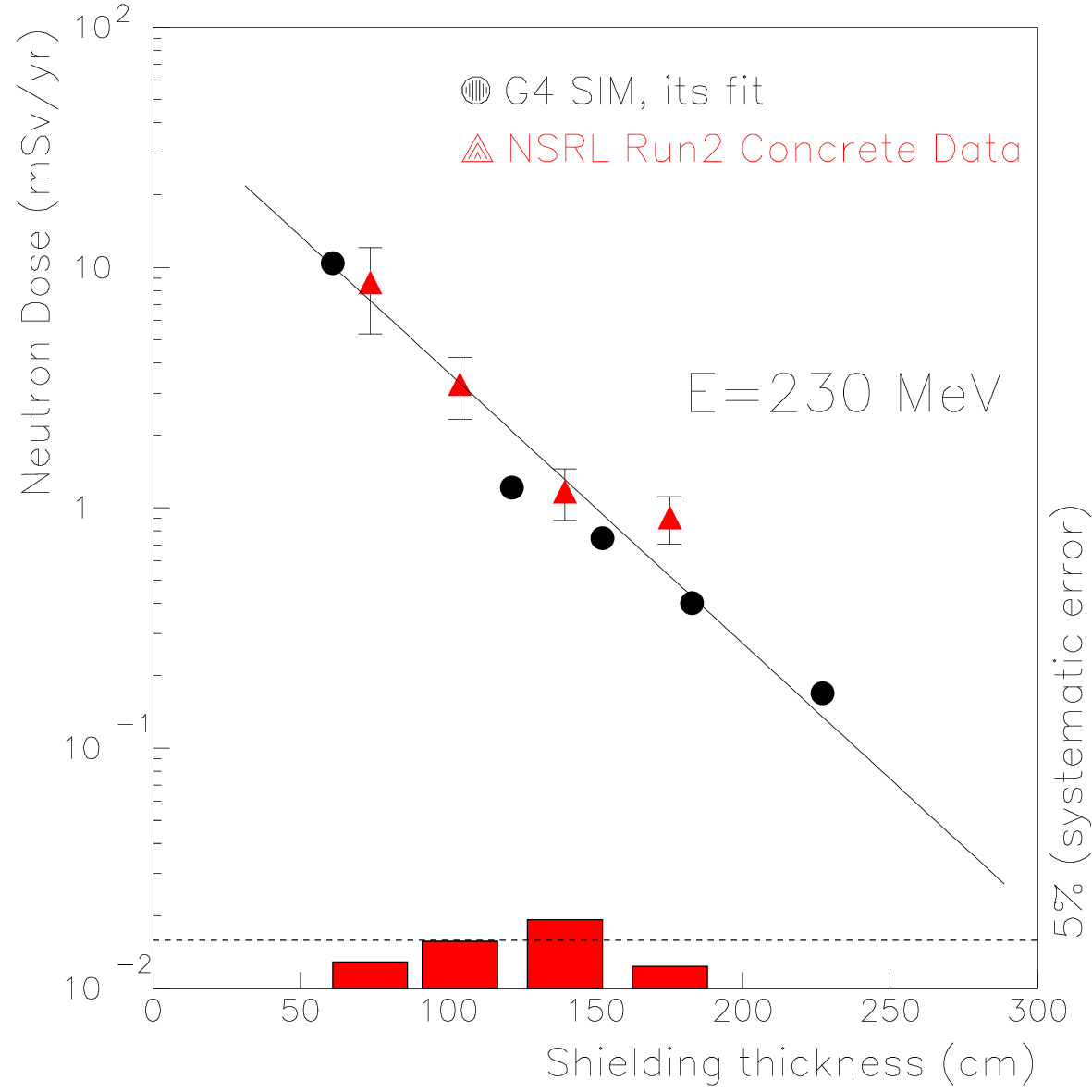}
    \caption[ComparisonSim2Data]{
(color online) The estimation of annual neutron dose as a function of shielding thickness (cm) between GEANT4 (Black circles) and NSRL-Run 2 (Red triangles) for proton energy E=150 MeV (left), E=190 MeV (middle), E=230 MeV (right). Solid line is a fit with an exponential function. Red bars on the x-axis are systematic uncertainty of the measurement. The horizontal dotted line is a 5\% of uncertainty index.
   }\label{fig:Sim2Data}
   \end{center}
 \end{figure*}

Figure~\ref{fig:Sim2Data} shows the annual neutron dose estimation assuming the $10^{14}$ protons in the workload per year. It shows both results from the simulation (black points) and NSRL-Run 2 data (red points) for the proton beam energies, E=150, 190, and 230 MeV as a function of shielding thickness.

\subsection{Neutron dose comparison with world data for concrete}
 We compared our GEANT4 simulation and NSRL-Run 2 experimental data with world data from S.~Agosteo {\it et al.}~\cite{SAgosteo}. This comparison focused on the annual neutron dose estimation (measured in $\mu$Sv/yr) as a function of shielding thickness for different proton beam energies.
 
\begin{figure}[!htb]
   \begin{center}
	\includegraphics[angle=0,width=75mm,height=80mm]{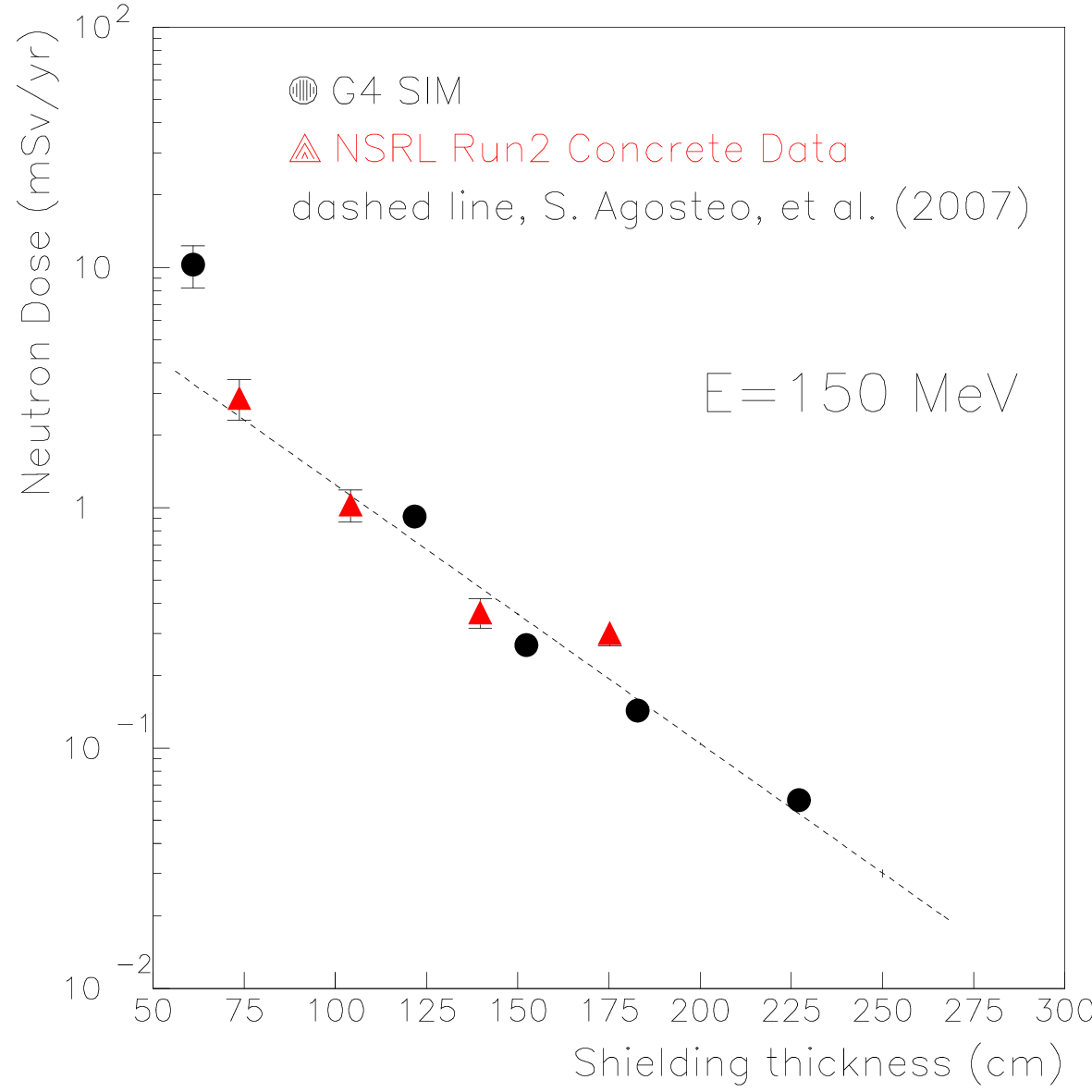}
        \caption[ComparisonWorldData]{
(color online) The estimation of annual neutron dose as a function of the concrete shielding thickness (cm) between GEANT4 (Black circles), NSRL-Run 2 (Red triangles), and world data (dashed line)~\cite{SAgosteo} for proton energy E=150 MeV.
   }\label{fig:DataCompSAgosteo}
   \end{center}
 \end{figure}

Figure~\ref{fig:DataCompSAgosteo} presents an example of this comparison, illustrating neutron dose results from NSRL-Run 2, GEANT4 simulations, and published world data. 
The reference~\cite{SAgosteo} dashed curve was obtained from the 150 MeV FLUKA concrete attenuation calculation reported by Agosteo et al. Their calculation considered secondary radiation generated when 150 MeV protons were stopped in a thick iron target and subsequently attenuated in ordinary concrete. The published attenuation data were digitized/converted to the annual-dose normalization used in the present work. Because the neutron-production target and source geometry differ from those of the present NSRL experiment, this comparison is used only as a reference for the magnitude and attenuation behavior rather than as a direct validation under identical conditions.

Once we obtained the neutron dose as a function of shielding thickness, we extracted the neutron linear attenuation coefficient (LAC, $\mu$) and the mass attenuation length (MAL, $\lambda_m$) from exponential fits. Both $\mu$ and $\lambda$ were determined from our simulation and experimental data and compared with published reference data. The attenuation length is given by $\lambda = 1/\mu$, while the mass attenuation coefficient is defined as $\mu/\rho$, where $\rho$ is the density of the shielding material. Therefore, mass attenuation length ($\lambda_m$) is $\rho/\mu_m$.

The neutron spectrum from our Monte Carlo simulations is dominated by low-energy contributions; therefore, we focused our comparison on reference data in this energy regime.
Table~\ref{tab:lac} shows the neutron LAC obtained for our concrete sample, compared with literature values for concrete, which are typically given as ranges due to density variations~\cite{RMultafiatin}. The MAL is also compared with the average values reported by the Particle Data Group (PDG)~\cite{PDG} for neutron energies below 20 MeV. The concrete density assumed in the PDG data is 2.4 g/cm$^3$, whereas our sample has an average density in the range of 2.2 – 2.4 g/cm$^3$.
For the data, the five-block point is associated with the largest attenuation and therefore the lowest measured dose. Consequently, its relative uncertainty is larger and its influence on the fitted attenuation coefficient is potentially significant. To evaluate this effect, the attenuation fit was repeated after excluding the five-block point. The resulting change in the fitted attenuation coefficient was 2.5\%, demonstrating that the reported attenuation behavior is robust to the final data point.

\begin{table}[!htb]
\centering%% 
\caption{Neutron linear attenuation coefficient ($\mu$) and mass attenuation length ($\lambda_m$) of concrete comparison to the world data}\label{tab:lac}
\renewcommand{\arraystretch}{1.2} 
\begin{tabular}{c|c|c}%%
\hline
Variable & World data & This work \\
\hline\hline
LAC, $\mu$ (cm$^{-1}$)
& $0.17$--$0.19$
& $0.08 \pm 0.03$ \\
MAL, $\lambda_m$ (g/cm$^2$)
& 30
& $28.5 \pm 0.8$ \\
\hline
\end{tabular}
\end{table}

\subsection{Simulation validation with measurement for the magnetite shielding}
In the previous section, we presented a comparison of our concrete neutron dose rate, linear attenuation coefficient, and attenuation length with GEANT4 simulation results and published world data. Overall, the comparison with published concrete data provides a consistent check of the present experimental and simulation methodology. The attenuation length obtained in this work is reasonably close to the published reference value; however, the fitted linear attenuation coefficient shows a larger difference. These differences may arise from differences in neutron source, shielding composition and density, experimental geometry, and fitting methodology. Therefore, the present comparison should be regarded as a consistency check rather than a definitive validation under identical conditions.

\begin{figure}[!htb]
   \begin{center}
	\includegraphics[angle=0,width=75mm,height=80mm]{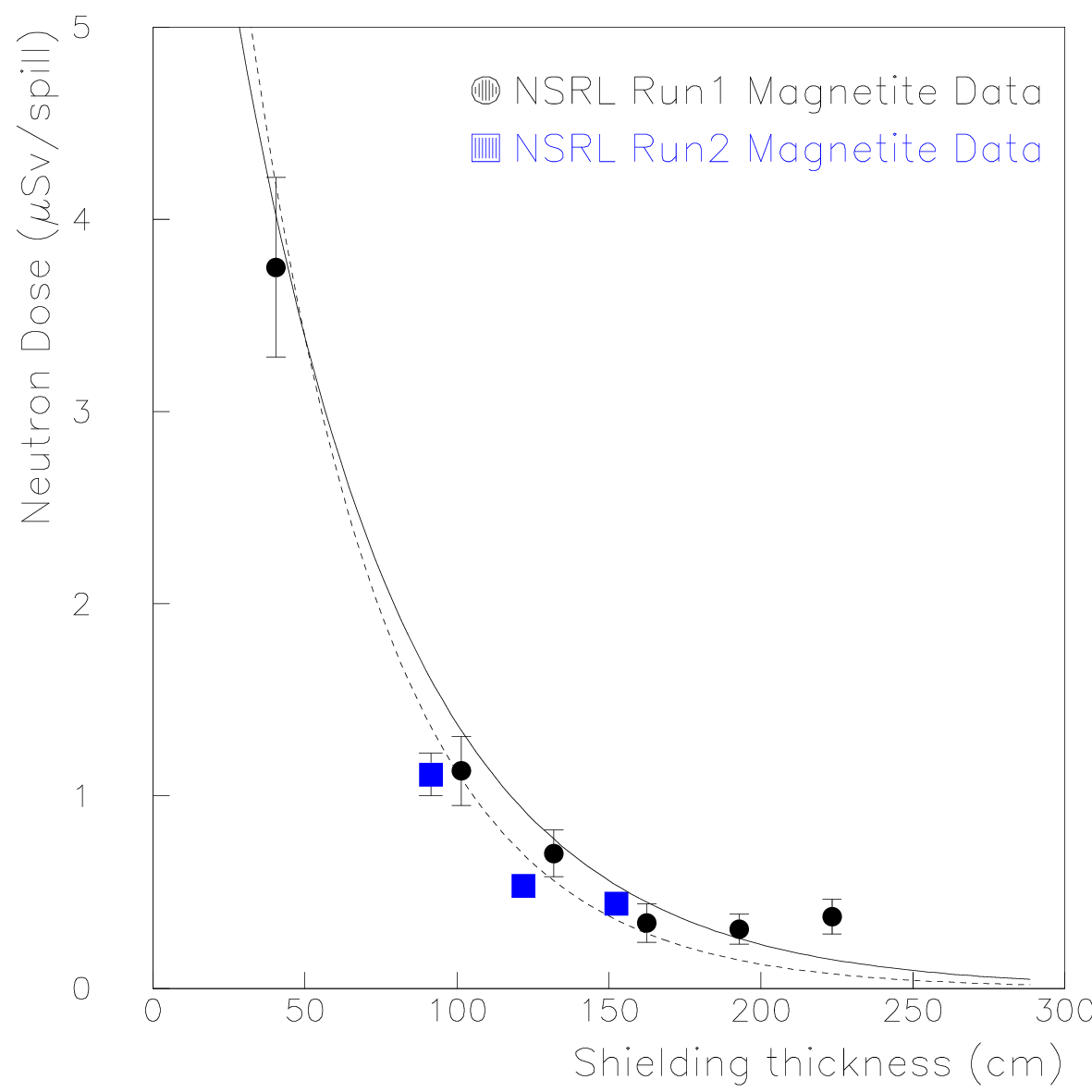}
        \caption[MagExpRun1n2]{
(color online) The neutron dose per proton spill ($\mu$Sv/spill) as a function of the shielding thickness of the magnetite material. The error bars are statistical uncertainties. Run 1 had 200 MeV and Run 2 with 190 MeV of the proton energy. Curves are fits of data with exponential functions.
   }\label{fig:magnetite_exp1_n_exp2}
   \end{center}
 \end{figure}

Building on this validation, we conducted a similar study using magnetite as the shielding material, incorporating both GEANT4 simulations and experimental measurements. The GEANT4 simulations and experimental measurements exhibit comparable attenuation trends for magnetite. Differences in the absolute dose values remain, and the results should be interpreted in the context of the estimated experimental and systematic uncertainties. However, magnetite demonstrated a significantly enhanced neutron shielding effect—approximately three times greater than that of concrete. Figure~\ref{fig:magnetite_exp1_n_exp2} presents the experimental results from NSRL-Run 1 and NSRL-Run 2, highlighting the consistent attenuation performance of magnetite.

Figure~\ref{fig:magnetite_exp1_n_exp2} shows a slight downward shift in dose in the Run 2 data. This shift is attributed to differences in shielding block positioning, primary proton beam energy between Run 1 and Run 2. An overall normalization factor was applied when comparing data acquired during Run 1 and Run 2 to account for the difference in the number of incident protons per measurement and the corresponding dose-per-spill normalization. The factor is a multiplicative normalization and does not modify the shielding-dependent attenuation shape.
To ensure a consistent comparison, we converted the measured neutron dose ($\mu$Sv/spill) for a 5-block thickness to an annual dose (mSv/yr) estimate, assuming a workload of $10^{14}$ protons. This converted measurement was then compared to the corresponding GEANT4 simulation result. 
Table~\ref{tab:magComp} presents the direct comparison of magnetite shielding material between simulation and experimental measurements, highlighting the consistency between both approaches.
Both the simulation and experimental measurements show consistency within our overall uncertainty of 17-36\%. The identical experimental setup and repeated measurements in Run 2 further verified the results from Run 1.\\

\begin{table}[!htb]
\centering%% 
\caption{Neutron doses with statistical and systematic uncertainty per proton beam spill and annual neutron dose estimation using the workload with the five magnetite blocks from the GEANT4 simulation and experimental results. Values are reported as dose $\pm$statistical uncertainty $\pm$systematic uncertainty.\\}\label{tab:magComp}
\renewcommand{\arraystretch}{1.2} 
\begin{tabular}{c|c|c}%%
\hline
  Method & Dose ($\mu$Sv/spill) & Dose (mSv/yr) \\ %% 
\hline
\hline
GEANT4    & ~0.29$\pm$0.03$\pm$0.03 &  ~0.14$\pm$0.02$\pm$0.02\\
NSRL-Run 1 & ~0.34$\pm$0.04$\pm$0.05 &  ~0.17$\pm$0.02$\pm$0.02\\
NSRL-Run 2 & ~0.44$\pm$0.05$\pm$0.06 &  ~0.22$\pm$0.03$\pm$0.03 \\
\hline
\end{tabular}
\end{table}
% Proton spill = 1~2x10^11,  Workload=2~4*10^14  for simulation

\section{Systematic Uncertainty}

We investigated the major sources of systematic uncertainty that could affect our results. The systematic uncertainty in the experimental data analysis was evaluated based on several variables, with the estimated values summarized in Table~\ref{tab:sysError}. 

The major systematic uncertainties in this analysis arise from the determination and subtraction of the background contribution in the signal detector. As explained earlier, the background extraction was performed using the background detector, which is located in the open area. The accuracy of the background determination depends heavily on the selection of the event angle in D1, which is crucial for removing the direct neutron contributions from the water tank. We used two different functional forms (exponential and second-order polynomial functions) to fit the transverse background events as a function of distance between two detectors, D1 and D2. 

The difference between the background subtractions obtained using these methods is quoted as the systematic uncertainty. The quoted uncertainty for the background identification and functional form for fit were found approximately 10\% and 5\% each. However, the difference in detection efficiency between the two detectors was found to be minimal ($\pm$0.5 \%). 

Another major source of systematic uncertainty is the determination of the shielding block density for the concrete. The blocks were constructed using a cubic-foot volume mold; however, the concrete shape became distorted during the drying process, resulting in blocks that were not perfectly cubic. To measure the concrete density accurately, the shielding blocks were divided into small sample pieces. The sample volumes were measured using the Archimedes method, and their masses were measured with a scale. Both weight and volume measurements were performed three times, and the mean values were used. Using this approach, the volume uncertainty was estimated to be approximately 1–6\%, which propagates directly into the density calculation. The density of each block was then determined from its mass and the averaged volume.

Reference~\cite{RGBaonza} evaluated the reliability of the detector response function with respect to the nuclear data library used in their analysis. The study concluded that the response function of the WENDI-II detector exhibited strong dependence on the nuclear data library in the high-energy neutron regime, particularly above 150 MeV. This dependence indicates that the dominant interaction mechanisms vary with neutron energy.
However, for neutron energies below 150 MeV, the dependence on the data library was found to be less than 3\%.
We quoted and added the uncertainty associated with the use of different nuclear data libraries in the analysis. This uncertainty is directly associated with the conversion factor used in the dose calculation.

Other potential sources of uncertainty, including shielding distance including air gaps ($<\pm2\%$), detector position ($<\pm1\%$), neutron detection efficiency, and primary proton beam energy ($\pm$ 0.7-1.2 \%), were considered but determined to be not in the major uncertainty.
We rule out gamma contamination in the WENDI-II detector by referencing its excellent gamma rejection factor ($<$ 10 cpm through 100 mSv/hr) within our energy regime~\cite{WENDI_LAN}.

\begin{table}[!htb]
\centering%% 
\caption{Various sources of systematic uncertainty and its estimated value.\\}\label{tab:sysError}
\renewcommand{\arraystretch}{1.2} 
\begin{tabular}{l|c}%%
\hline
 Source of Uncertainty   & Relative Dose Difference  \\ %% 
\hline
\hline
BG (L/T) identification           &  4-11 \% \\
BG functional form                &  5 \% \\
Concrete density                  &  1-6 \% \\
Conversion coefficient in model &  3 \% \\
\hline
Total &  7.1-13.8 \%\\
\hline
\end{tabular}
\end{table}

\section{Concrete and Magnetite Comparison}

Since we obtained neutron dose measurements for both magnetite and concrete under the same shielding configurations, we evaluated the relative neutron shielding effectiveness of magnetite by calculating the neutron dose ratio between the two materials.
This ratio ($R_{M/C}$) was determined across various proton energy levels, using a consistent shielding thickness of 1.524 m (or 5 blocks). Table~\ref{tab:ratio} presents the annual neutron dose estimates for magnetite and concrete shielding barriers, derived from both experimental data and GEANT4 simulations.

The principal comparison in this work is performed at equal geometric thickness because shielding-space requirements are a primary consideration in facility design. However, because magnetite and concrete have substantially different densities, equal geometric thickness does not correspond to equal mass thickness. We therefore additionally evaluate the materials using mass thickness. This comparison provides a complementary measure of attenuation performance and avoids attributing the entire observed difference to geometric thickness alone.

\begin{table}[!htb]
\centering%% 
\caption{The annual neutron dose and dose ratio ($R_{M/C}$) between magnetite and concrete for three proton beam energies with five blocks shielding thickness from measurements and simulations.\\}\label{tab:ratio}
\renewcommand{\arraystretch}{1.2} 
\begin{tabular}{c|c|c|c}%%
\hline
  Proton E & ~Aggregate~ & ~EXP (5-block)~ & ~SIM (5-block)~ \\ 
\hline
\hline
150 MeV & Magnetite &   0.10 mSv/yr& 0.06 mSv/yr\\
150 MeV & Concrete &   0.30 mSv/yr& 0.19 mSv/yr\\
        & $R_{M/C}$ & 0.33  & 0.32 \\
\hline   
190 MeV& Magnetite &   0.17 mSv/yr& 0.14 mSv/yr\\
190 MeV& Concrete &   0.59 mSv/yr& 0.47 mSv/yr\\
    & $R_{M/C}$ & 0.29  & 0.30 \\
\hline
230 MeV& Magnetite &   0.25 mSv/yr& 0.15 mSv/yr\\
230 MeV& Concrete &   0.91 mSv/yr& 0.52 mSv/yr\\  
    & $R_{M/C}$ & 0.27  & 0.29 \\
\hline
\end{tabular}
\end{table}

 \begin{figure}[!htb]
   \begin{center}
	\includegraphics[angle=0,width=75mm,height=80mm]{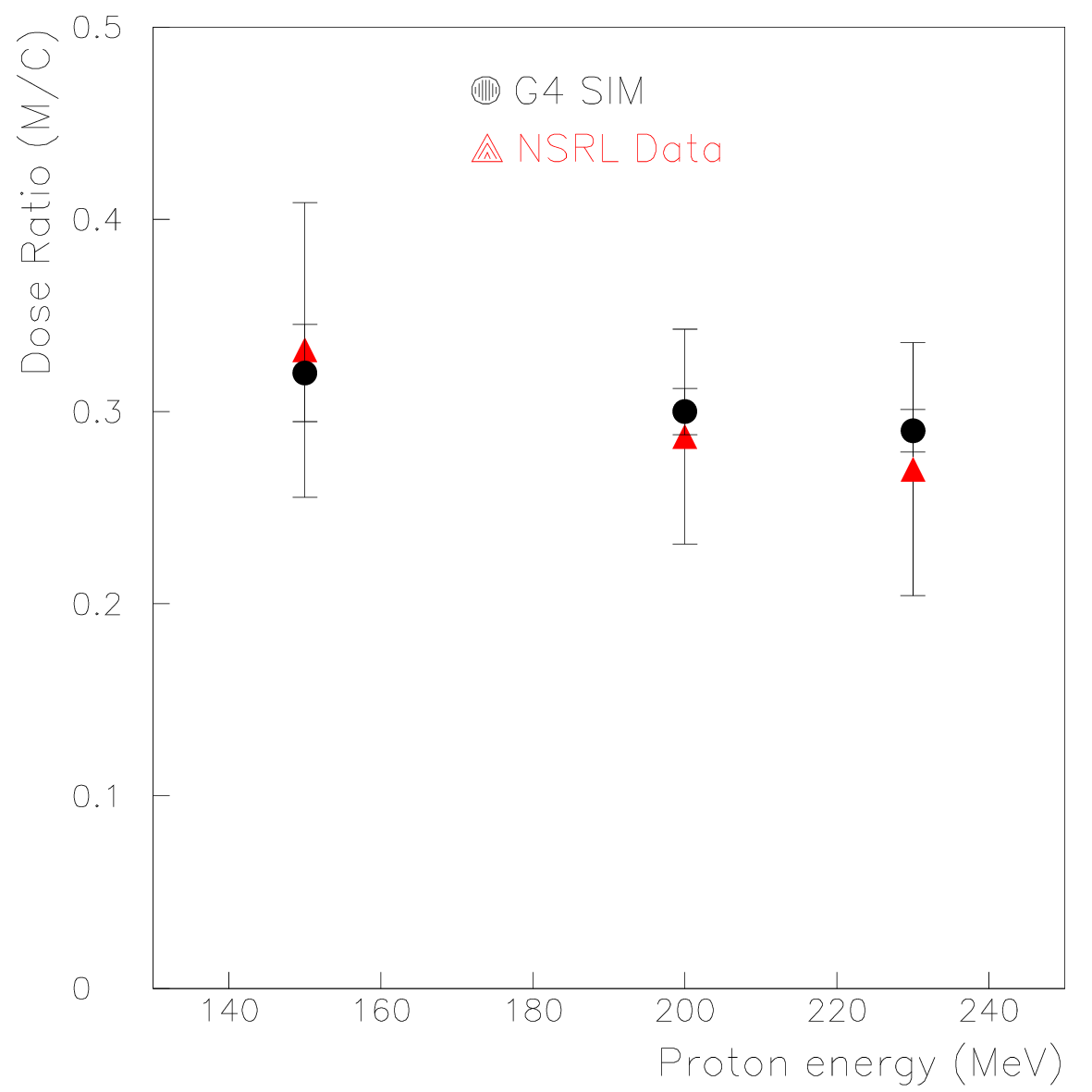}
        \caption[MagExpRun1n2]{
(color online) The neutron dose ratio, magnetite over concrete ($R_{M/C}$), as a function of proton beam energy from the experimental data (red triangles with outer error bars) and GEANT4 simulation (black circles with inner error bars). 
   }\label{fig:ratio}
   \end{center}
 \end{figure}
 
Table~\ref{tab:ratio} also presents the dose ratio between magnetite and concrete for all three proton energies, based on both experimental measurements and GEANT4 simulations. 
Figure~\ref{fig:ratio} illustrates these ratios as a function of proton energy.
The results indicate no significant energy dependence in the shielding effectiveness. 
For the investigated five-block configuration, the measured magnetite-to-concrete dose ratios were 0.27–0.33, corresponding to approximately a threefold lower neutron dose for magnetite at the same nominal geometric thickness.

We monitored the activation of the shielding blocks after the measurements. Table~\ref{tab:activation} summarizes the activation history after experiments for both concrete and magnetite blocks. In 2023, the measurement was done with only magnetite. We surveyed all eight shielding blocks immediately after completing the measurements. The survey meter indicated elevated activation (“hot”) for the first block placed directly behind the water tank, while the remaining blocks showed activation levels that were very low and not recordable.

In 2024, we repeated the experiment with approximately twice the beam time used in 2023. In this time, our measurement was focused on the concrete and magnetite. The activation was checked again six months later. At that time, the activation levels for both the magnetite and concrete blocks were still very low, though slightly above background.

\begin{table}[!htb]
\centering%% 
\caption{The record history of the activation level for concrete and magnetite. The recording was made for the highest activation level which the block is located the closest to the water tank. See Fig. \ref{fig:setup}.\\} \label{tab:activation}
\renewcommand{\arraystretch}{1.2} 
\begin{tabular}{c|c|c|c|c}%%
\hline
  Measured & Exposed & Survey & Magnetite & Concrete  \\ 
%\hline
 Date &  Hour & Date & Dose($\mu$R/hr) & Dose($\mu$R/hr)\\ 
\hline
\hline
   8/14/2023 & 8 &   &  &   \\
   NSRL-Run 1 &  &   &  &   \\
\hline
&  & 8/15/2023  & 150  & \\
    &  & 8/18/2023  & 42  &    \\
    &  & 8/25/2023  & 16 &    \\
    &  & 8/31/2023  & 12 &    \\
\hline
   9/6/2024 & 16 &   &  &   \\
   NSRL-Run 2 &  &   &  &   \\
\hline
    &  & 3/27/2025  & 6 & 8  \\
    &  & 6/16/2025 & 5  &  5  \\
    &  & 9/12/2025  &   &  9-10   \\
\hline
\end{tabular}
\end{table}

\section{Summary}
 In summary, we report on the neutron shielding characteristics of both concrete and magnetite aggregate by comparing experimental results with simulation calculations. The experimental and GEANT4 results exhibit comparable attenuation trends for both concrete and magnetite, although differences in absolute dose remain between measurement and simulation. Our findings indicate that magnetite exhibits superior neutron shielding performance compared to concrete in terms of total neutron dose, linear attenuation coefficient, and attenuation length for the same barrier thickness for the incident proton energies measured. Because magnetite and concrete have substantially different densities, this comparison is specific to the investigated geometric thickness and should not be interpreted as a density-independent intrinsic shielding factor.

Beyond its superior neutron shielding properties, magnetite also offers a significant advantage in construction efficiency when used as infill to shielding modules. The integration of modular building technology in radiation facility construction can substantially reduce the overall construction period, enabling faster facility completion and earlier revenue generation. It can also be removed more effectively should a building need to be accomplished. This represents a novel approach in radiation shielding infrastructure.
Compared to traditional concrete-based shielding, magnetite requires a smaller volume to achieve the same radiation attenuation, making it a more space-efficient solution.

We sincerely acknowledge the exceptional efforts of the staff at the Accelerator and Physics Division of Brookhaven National Laboratory, whose dedication and expertise made these experiments possible.

\end{document}